# Effects of Mercury surface temperature on the sodium abundance in its exosphere

**E. Rognini**[1,2], A. Mura[3], M. T. Capria[3], A. Milillo[3], A. Zinzi[1], V. Galluzzi[3]

**Corresponding author**: Edoardo Rognini, Agenzia Spaziale Italiana (ASI), Via del Politecnico snc, Rome, Italy. Email: *edoardo.rognini@ssdc.asi.it*

[1] ASI-SSDC, via del Politecnico snc, I-00133 Rome, Italy

[2] INAF-OAR Osservatorio Astronomico di Roma, Via Frascati 33, 00040, Monte Porzio Catone (RM), Italy

[3] INAF-IAPS, via Fosso del Cavaliere 100, I-00133 Rome, Italy

Key points

- The link between the surface temperature of Mercury and the exosphere sodium abundance has been analyzed.
- A thermophysical model is being used coupled with an exospheric model.
- The exospheric sodium abundance and distribution is strongly dependent on the surface thermal map

**Abstract**

The link between the surface temperature of Mercury and the exosphere sodium content has been investigated. Observations show that, along the orbit of Mercury, two maxima of total Na content are present: one at aphelion and one at perihelion. Previous models, based on a simple thermal map, were not able to reproduce the aphelion peak. Here we introduce a new thermophysical model giving soil temperatures as an input for the IAPS exospheric model already used in the past with the input of a simple thermal map. By comparing the reference model output with the new one, we show that such improved surface temperature map is crucial to explain the temporal variability of Sodium along the orbit.

# 1. Introduction and background

Ground-based and MESSENGER (MErcury Surface, Space ENvironment, GEochemistry and Ranging) observations indicate that the composition of the constituent particles in the Mercury's environment includes, H and He, Na and $Na^+$, K, Mg, Ca and $Ca^+$, Mn, Fe and Al (Broadfoot et al. 1974; Potter and Morgan 1985, 1986; Bida et al. 2000; McClintock et al. 2008; Bida and Killen 2017). Many other species and molecules are expected in neutral or ion form. The outermost part of Mercury's environment, its exosphere, is known to be the effect of different surface release processes, acting differently for different exospheric species. One of the most abundant species and surely the most observed one from space and from ground is Sodium. This alkali also demonstrated a peculiar behavior of spatial distribution, short (e.g.: Massetti et al 2017; Mangano et al 2015; Leblanc et. al. 2009; Killen et al 2001; Potter et al. 1999) and long term variability (Cassidy et al 2016; Milillo et al. 2021; Potter et al 2006)

The surface temperature and its variation have a strong influence on the Na exospheric generation. The MESSENGER mission observed the exospheric sodium emission for more than 16 Mercury years with the MASCS instrument (Mercury Atmospheric and Surface composition Spectrometer); the sodium emission as a function of time has been characterized, and a pattern varying over the Hermean year has been observed [Cassidy et al., 2016]. The projection of the sodium column density on the equatorial plane reveals an enhancement at -90° and +90° longitude when they are illuminated by the Sun; these longitudes are called "cold poles" because of their lower-than-average temperature allowing a preferential condensation of Na atoms that re-impact the surface from the exosphere (Leblanc and Johnson, 2010). This can be considered as a strong indication of the surface temperature effects on the exosphere, and it has been also partially confirmed by the data set obtained by ground based observation of THEMIS solar telescope (only when the "cold pole" is at dawn, see Milillo et al 2021). However, some unexplained features are present (e.g. the behavior of southern and northern peaks along the year in relation of local time), so it is important to use models in order to interpret the data in preparation for the new BepiColombo mission.

BepiColombo is the first European space mission to Mercury. The spacecraft have been launched in October 2018, and will reach its destination in December 2025; the surface, the exosphere and the magnetosphere of the planet will be studied in detail. The mission comprises two spacecrafts: the Mercury Planetary Orbiter (MPO) and the Mercury Magnetospheric Orbiter (MMO). The orbit of the two spacecrafts and their scientific payload "will allow a wider range of scientific questions to be addressed than those that could be achieved by the individual instruments acting alone, or by previous missions" (Milillo et al., 2020). The MPO module is equipped with MERTIS (Mercury Radiometer and Thermal Infrared Spectrometer), an IR imaging spectrometer that will provide detailed information on the surface's mineralogical composition and temperature. The SERENA-STROFIO instrument (STart from a ROtating FIeld mass spectrOmeter (Orsini et al. 2021) onboard the BepiColombo mission will measure exospheric abundances with a high resolution. The data obtained with these instruments will be a crucial test for the existing thermophysical and exospheric models.

In particular, the spatial distribution of Mercury's sodium abundance is a key parameter for understanding the processes in the exosphere of the planet, and their link with the surface temperature. Waiting for the more detailed data of BepiColombo mission, to describe the sodium abundance and its evolution during the year we coupled a thermophysical model (section 2.1) with the IAPS model of the exosphere (section 2.2): the output of the thermophysical model, the surface temperature of the Hermean soil along the orbit at different latitudes, is used as input for the exospheric circulation model. The results of their application are described and discussed in section 3. Conclusions are given in Section 4.

## 2. The models

### 2.1. The thermophysical model

We used a one-dimensional model [*Capria et al.*, 2014, *Rognini et al.*, 2019] that solves the heat conduction equation and provides theoretical temperatures as a function of thermal conductivity, density, specific heat, albedo, and emissivity. Thermal inertia is defined as

$$I = \sqrt{\rho k c} \quad (1)$$

where $\rho$ is the density, $k$ the thermal conductivity and $c$ the specific heat. Thermal inertia is a fundamental indicator of the physical properties of a soil, and provides information on the age and geological history of the surface of airless bodies.

The 1D heat transport equation is solved with a finite-differences scheme applied to each of the layers into which the internal radius of the body has been subdivided:

$$\rho c \frac{\partial T}{\partial t} = \frac{\partial}{\partial x}\left(k \frac{\partial T}{\partial x}\right) \quad (2)$$

where $T$ is the temperature, $t$ the time, and $x$ the depth. The thickness of the layers is 0.05 cm close to the surface and is increasing with depth. The surface boundary condition is written as follows:

$$\frac{S f_{sun}(1-A)\mu}{r^2} = X\sigma T^4 - k\frac{\partial T}{\partial x} \quad (3)$$

where $S$ is the solar constant ($1361\ W\ m^{-2}$), $f_{sun}$ is a factor taking into account the finite angular size of the Sun, $A$ is the Bond albedo, $\mu$ is the cosine of the illumination angle, $r$ is the heliocentric distance in AU, $\sigma$ is the Stefan-Boltzmann constant, and $X = (1 - \varepsilon\xi)\varepsilon$ is a "roughness parameter" where $\varepsilon$ is the emissivity and $\xi$ is the sub-pixel roughness. This parameter [*Lagerros*, 1998] has been introduced to characterize the topography (roughness) on sub-pixel scale; it can be regarded as a measure of the surface irregularity at a scale smaller than the shape model and larger than the thermal skin depth (of the order of 1 cm, expressed, in the case of the diurnal cycle, as $D = \sqrt{\frac{kP}{\pi \rho C}}$, where $P$ is the rotation period); the roughness can be interpreted, for example, as the percentage of cratered terrain with respect to flat terrain [*Müller and Lagerros*, 1998; *Keihm et al.*, 2012]. Roughness [*Davidsson et al.*, 2015] is thought to be the cause of thermal-infrared beaming, the phenomenon for which a surface emits in a non-lambertian way with a tendency to reradiate the absorbed radiation towards the Sun. The lateral heat conduction is neglected because the facets of the shape model on which the thermophysical model is applied are always larger than the thermal skin depth. The instantaneous illumination is taken into account by using a shape model of Mercury with a spatial resolution of 1°; the illumination angle, for each time step and location on the shape model, is obtained through SPICE-based software and navigational databases, the so-called "SPICE kernels" [*Acton*, 1996]. Anyway, a simple spherical shape for the planet can be also assumed as a good approximation.

We assume that the surface of the body is covered by a meters-thick layer of particulate material whose density is minimum on the surface and increases with the depth (see **Tab. 1**), following an empirical exponential law. The formulae for density and thermal conductivity are based on the properties of the particulate material, which have been derived from ground and spacecraft observations, lunar in-situ measurements and returned samples [*Capria et al.*, 2014; *Vasavada and Paige*, 1999; *Vasavada et al.*, 2012]. The thermal conductivity has the form:

$$k(T) = k_s + k_r T^3 \qquad (4)$$

where $k_s$ and $k_r$ are the conductive and radiative terms; the first term takes into account the conductive transfer across the particles, and the second term takes into account the radiative heat transfer, sensible at high temperatures. Different kinds of particulate material, ranging from very fine dust to incoherent rocky debris (regolith), can be simulated in the code. To each type of material a density profile, with values increasing with depth, has been associated. Thermal conductivity is taken into account through empirical expressions ($k_s$ and $k_r$) depending on density and temperature and based on the properties of the chosen particulate material, which have been derived from ground and spacecraft observations, lunar in-situ measurements and lunar returned samples [*Capria et al.,* 2014; *Vasavada and Paige* 1999; *Vasavada et al.* 2012; *Cremers and Hsia* 1973]. Laboratory experiments show that thermal conductivity depends on composition, particle size, porosity, and temperature [*Sakatani et al.*, 2017]. The range on thermal conductivity values induced by the composition is usually small with respect to the range induced by the changes on physical structure of the regolith particle assemblage; the thermal conductivity (and then thermal inertia) is then more a function of the physical structure of the soil than the composition, so the expressions used in *Rognini et al.* (2019) can also be used for the surface of Mercury. The expressions used to evaluate the thermal conductivity are derived from Cremers paper (and figures) on Apollo lunar data as described and used in a technical ESA report on Mercury environmental specifications (SCI-PF/BC/TN01), written as part of the BepiColombo mission definition study. In that report Cremers work is used to derive empirical expressions that can be used to describe the thermal conductivity of a particulate soil with density slightly increasing with depth. Typical thermal conductivity values range from about $10^{-3}$ (fine dust) to 0.02 Wm$^{-1}$K$^{-1}$ (regolith) at the temperature range of the simulations.

Specific heat is temperature dependent and is derived from a fit of experimental measurements [*Biele et al.* 2014]:

$$c(T) = a_0 + a_1(T - T_0) + a_2(T - T_0)^2 + a_3(T - T_0)^3 \qquad (5)$$

where $T_0 = 250$ K, $a_0 = 633$ J kg$^{-1}$ K$^{-1}$, $a_1 = 2.513$ J kg$^{-1}$ K$^{-2}$, $a_2 = -0.0022$ J kg$^{-1}$ K$^{-3}$, $a_3 = -2.8 \times 10^{-6}$ J kg$^{-1}$ K$^{-4}$.

The normal albedo (the albedo at the zenith of the surface element) and emissivity are set to $A_0=0.12$ (BepiColombo report SCI-PF/BC/TN01) and 0.82 respectively. The albedo at illumination angle $\theta$ is calculated by using the relationship

$$A(\theta) = A_0 + a \left(\frac{\theta}{\pi/4}\right)^3 + b \left(\frac{\theta}{\pi/2}\right)^8 \qquad (6)$$

with $a = 0.06, b = 0.25$ [Hayne et al. 2017].

| Material | Minimum density (kg/m³) | Maximum density (kg/m³) | Thermal inertia (Jm$^{-2}$ s$^{-½}$K$^{-1}$) |
|---|---|---|---|
| Fine dust | 1200 | 1800 | 1-16 |
| Fine regolith | 1350 | 1950 | 50-65 |
| Regolith | 1350 | 1950 | 110-140 |

**Table 1.** Density ranges for the materials considered in this work. The hypothesis is made that density is increasing with depth: the minimum density is found on the surface.

As already stated, the angular size of the Sun is taken into account by including the $f_{sun}$ factor (**eq. 3**), calculated by the fraction of visible solar disk and the limb darkening [*Glaser et al*. 2019]. The intensity $I(\alpha)$ observed at angle $\alpha$ away from the center is given by the limb-darkening formula

$$I(\alpha) = I(0)\{1 + \sum_{k=1}^{2} W_k (1 - \cos \varphi)^k\} \qquad (7)$$

where $\varphi$ is the emission angle and $W_1 = -0.47, W_2 = 0.23$ [Cox 2000]. The intensity is wavelength-dependent, in this case the wavelength is fixed at 550 nm (center of the solar spectrum). When Mercury is at a distance $r$ from the Sun the visible solar disk radius is

$$R_{vis} = R_\odot \cos \gamma \qquad (8)$$

where $R_\odot$ is the radius of the Sun and $\gamma$ is the angular radius of the Sun, given by $\sin \gamma = R_\odot/r$. The visible solar disk is divided into a set of $N$ concentric rings, and for every ring the visible area is calculated:

$$A_j = \pi(R_{j+1}^2 - R_j^2)f_j \qquad (9)$$

where $R_j$ and $R_{j+1}$ are the boundary of the ring $j$ and $f_j$ is the fraction of the visible ring area. The total solar irradiation is proportional to the sum

$$I_{tot} = \sum_{j=1}^{N} A_j \left\{ 1 + \sum_{k=1}^{2} W_k (1 - \cos \varphi_j)^k \right\} \qquad (10)$$

and the solar factor is calculated with the formula

$$f_{sun} = \frac{I_{tot}}{I_{max}} \qquad (11)$$

where $I_{max}$ is the total intensity obtained when the entire solar disk is visible ($P_j = 1$ for every ring). The effect of the finite solar angular dimension is important at sunrise and sunset, and for high latitudes (within 1.5 degrees from the poles).

When available, observed temperatures can be compared with the theoretical temperature calculated with the thermophysical model, and the thermal properties of surface and subsurface (as thermal inertia) can be retrieved. The temperature curve (temperature as function of time) may indicate discontinuities in parameters (thermal conductivity, specific heat etc.) in the sub-soil regions due to physical heterogeneity; this effect is particularly important at nighttime, and at dawn and sunset due to the quick changes in temperature.

## 2.2. The exospheric model

The model for the calculation of exospheric quantities is a MonteCarlo (MC) single particle model and, at Mercury, it has been used extensively to study the sodium distribution and circulation [Mura et al., 2005, 2006a, 2006b, 2007, 2008, 2009, 2012].

The observed overall Na exosphere is compatible with an energy distribution and an intensity resulting from Photon-stimulated Desorption (PSD, Mura et al., 2009) and/or Thermal desorption (TD, Gamborino et al. 2020). But, as we describe below, the spatial and temporal distribution cannot be explained invoking only these processes, so a more complicated modelling is needed, including mutual effects between different processes. While the micrometeoroid impact vaporization is a stochiometric release process that depends mainly on meteoroid velocity, and that hardly explain the observed double peaked spatial distribution and the time variability of the Na exosphere, it seems the main surface release process for refractories peaking mainly at the dawn terminator (e.g. Milillo et al 2020; Grava et al. 2021). Ion Sputtering (IS) seems a process that could better be related to the observed distributions and short-term variabilities linked to fluctuation of the impacting solar wind onto the surface. But the intensity of the observed exosphere is hardly justified by this single process, which generally has a low release efficiency. For example, observations by Schleicher et al., 2004 can't be reproduced by using IS, neither in density, nor in scale height or velocity distribution (Mura et al., 2009). Electron stimulated desorption could occur, but we expect a lower efficiency in the dayside in nominal solar wind conditions, with respect to the PSD (Killen et al 2007). Probably this could be a relevant surface release process during strong solar events or in the nightside. Finally, the PSD and TD release efficiencies could be affected by other mechanisms onto the surface caused by external drivers, as described below. For this reason, the model considers as main surface Na release processes, PDS and TD, including possible effects by other conditions, like proton precipitation, steaking and surface condensation. In this study, we investigate the effect of the temperature on the global Na exosphere.

These surface release processes deplete the sodium abundance in the surface, which is refueled by diffusion from the interior of regolith grains (Killen et al 2004). The frequent observation of sodium exosphere distribution with double peaks at mid latitudes that roughly reflects the regions where the plasma is expected to precipitate [Schleicher et al., 2004] led to the hypothesis that solar wind precipitation enhances the diffusion inside the grains. Such enhanced diffusion is observed, for example, at the Moon [Sarantos et al., 2008]. Alternatively, it has been suggested that energetic ions (solar wind or magnetospheric ions) impact on the surface and induce chemical alteration (Mura et al. 2009). Sodium present in the rocks is liberated by proton chemical sputtering as described by the reaction:

$$2H + Na_2SiO_3 \rightarrow 2Na + SiO_2 + H_2O \qquad (12)$$

The ion implantation is close to the uppermost surface layer, so the Na atoms are produced in the uppermost surface and will thermally diffuse; then the sodium can be released into the exosphere by TD or PSD. Regardless whether the physical process is chemical sputtering or enhanced diffusion, the actual effect is that plasma precipitation gives "fresh" sodium particles to the surface, which can later be desorbed by PSD or TD. In addition, ejected sodium neutrals possibly precipitate onto different surface regions, thus varying the sodium abundance locally. In summary, a global model for the exosphere of Mercury (and, generally, for all species) should take into account the planetary rotation and orbit, the surface temperature, the surface chemical abundance, the exospheric circulation, and the precipitation of plasma. The model in this study is assuming that:

1. Since we are just comparing the global exospheric intensity along the Mercury orbit for two different temperature maps, we decided to simplify our model and use a single IMF orientation, with solar wind precipitating flux inversely proportional to the square of Sun-

Mercury distance ~~resulting~~. The assumed solar wind proton density at 1AU is 6 cm$^{-3}$, the velocity is 450 km/s and the constant IMF is (-15, 10, 10) nT, that is a realistic case that occurred during MESSENGER M1 flyby. A negative IMF$_x$ component and a northward IMF$_z$ lead to a preferential precipitation northward (Massetti et al 2007);, this correspond to moderate reconnection rate (Slavin et al. 2008). It is quite obvious that such IMF condition is not representative of a whole Mercury year. However, the IMF variability is not correlated with Mercury TAA, so that on average it does not explain the Na yearly variability; the only effect that cannot be neglected is the variability of the SW density with the Sun distance, which is, in fact, included. Short time variations of IMF could result in rapid changes of the proton precipitation, and produce Na exosphere short time variability, but this is not the subject of this study.

2. The sodium surface composition is calculated by taking into account the solar wind proton flux described in point 1, the planetary orbit and rotation, the thermal and photon-stimulated desorption, and the probability of photoionization of neutrals in their ballistic orbits in the exosphere. The surface composition is modelled with a regular 2D grid; the composition, function of time, is obtained with an ODE (Ordinary Differential Equations) solver. The migration of atoms from a surface element to the nearby ones is physically possible because exospheric atoms can precipitate far from their emission region. This is taken into account by including a flux term $\Phi_E$ that links each surface cell with all the others. Since the particles can leave a surface element at some time *t* and precipitate onto another element several time steps later, particles are temporarily stored in a buffer which has the physical meaning of the exosphere. The total number of particles in the buffer at a given time is, hence, the total contents of Mercury exosphere (but without any information on the distribution of such exosphere).

3. To calculate the detailed, 3D exospheric density at a given time, the sodium surface composition from point 2 is the input for the 3D Monte Carlo exosphere circulation model (Mura et al., 2009). This is needed because, in step 2, only the surface spatial and temporal variability is reconstructed. The model only calculates the start and stop point for exospheric particles ($\Phi_E$ in eq. 13) but does not retain the full trajectory to save computational memory.

### 2.2.1. Surface composition model

The planetary surface is divided into a set of surface elements (24°x48° latitude and longitude respectively), and for each surface element the sodium concentration as function of time $Y(t)$ is calculated by solving the differential equation:

$$N\frac{Y(t)}{dt} = q\Phi_{PREC} + \Phi_E - \Phi_{PSD} - \Phi_{TD} \qquad (13)$$

where $q$ is the coefficient for enhanced diffusion (10%), $\Phi_{PREC}$ is the flux of precipitating protons, $\Phi_E$ is the precipitating flux of exospheric particles as described in the previous section, $\Phi_{TD}$ and $\Phi_{PSD}$ are the TD and PSD local fluxes. As already mentioned, ϕE takes into account the atoms that fall back onto the surface, so that they are available for further desorption if they stick. The sticking probability is assumed as in Leblanc and Doressoundiram (2011) and it is function of the temperature. The fraction that does not stick to the surface is just bounced in thermal equilibrium with the surface - hence, is practically identical to a thermal desorbed population. At the end of this section we describe how ϕE is computed.

For PSD we use the formula:

$$\Phi_{PSD} = NY(t)\Phi_\gamma \sigma \cos(\alpha) \qquad (14)$$

where $N$ is the surface density ($7.5 \times 10^{14}$ cm$^{-2}$, Killen et al. 2001), $\Phi_\gamma$ is the relevant photon flux ($3 \times 10^{15}$ cm$^{-2}$s at 1 AU, and scaled with the distance from the Sun), $\sigma$ the PSD cross-section ($2 \times 10^{-20}$ cm$^2$, Yakshinskiy and Madey, 1999), $\alpha$ the angle from the subsolar point.

For TD, we use the formula:

$$\Phi_{TD} = \nu N C(t) e^{-U_{TD}/k_B T} \qquad (15)$$

where $\nu$ is the vibration frequency of adsorbed atoms ($10^{13}$ s$^{-1}$, Hunten et al. 1988) and $U_{TD}$ is the binding energy. The value of $U_{TD}$ has a strong effect on the desorption rate, and is comprised between

1.4 and 2.7 eV [Yakshinskiy et al. 2000]; an average value of 1.85 eV is used in this case, following Leblanc and Johnson (2003).

When a surface element is on the dayside, the concentration reaches its equilibrium value on a time scale $T_s = (\sigma \Phi_\gamma)^{-1}$; the parameters $\sigma$ and $\Phi_\gamma$ therefore mainly affect the equilibration time and not the equilibrium results. On nighttime, $C(t)$ increases due to the double contribution of the Na accumulation due to precipitating flux $\Phi_E$ of exospheric particles and of the plasma night side precipitation, $\Phi_{PREC} > 0$, until the surface element reaches the dawn terminator, then rapidly decreases due to the PSD contribution; a dawn-dusk asymmetry is therefore expected.

The particles released for TD usually have a velocity much lower than the escape velocity, so they always fall down onto the surface and TD does not contribute to the escaping flux from the surface. However, the TD causes a smearing of the places of Na release on the dayside, because the Na particles fall down within an area of few hundred kilometers.

Calculating the precipitating flux $\Phi_E$ of exospheric particles with high accuracy demands a considerably amount of resources. In, fact, the ballistic trajectories are deformed by the radiation pressure, they cannot be evaluated in an analytical way, and they require a numerical integration of motion. In steady state models such as Mura et al., 2005, this is done for a specific time. However, in this time-evolving model, it should have been done for each step (90 s), which is not computationally feasible. Instead, for simplicity, the flux is calculated with an ad-hoc Montecarlo model that launches several test particles with simple ballistic orbits (neglecting the radiation pressure) and evaluate (and retains in memory) only the precipitation location and travel time. The effect of radiation pressure is taken into account by decreasing the flux $\Phi_E$ by a factor that takes into account the number of Na particles that are lost in space at a given TAA, assuming the tables in Mura et al., 2011).

The fluxes are those of TD and PSD described above; the velocity distributions for this simple MC model are those given in Mura et al. [2009] and summarized below.

### 2.2.2. Exosphere circulation model

Once the surface sodium abundance has been calculated, the flux of particles leaving the surface is distributed along a large number of test particles, whose ballistic trajectories are simulated; the surface velocity distribution is a Maxwell-Boltzmann distribution, and the temperature is calculated with the thermophysical model as in section 2.1. A large number ($> 10^6$) of test sodium particles is simulated; for each particle the trajectory is calculated by taking into account the gravity and the radiation pressure acceleration; this latter can be up to 54% of the surface gravity in the case of sodium, ranging between 20 and 200 cm s$^{-2}$, and depends on the photon flux and the amount of Doppler shift out of the Fraunhofer features [Smyth and Marconi, 1995; Potter et al., 2002]. The radiative acceleration is calculated as in Leblanc and Johnson [2003], that is the Doppler shift is calculated as function of the velocity of each particle and from this the radiative acceleration is computed. The energy distribution of the emitted sodium atoms was extrapolated [Johnson et al. 2002] by laboratory measurements of electron stimulated desorption of adsorbed Na from amorphous ice that is similar to the energy distribution for PSD; Johnson et al. [2002] found a good analytical function approximating the spectrum that is given by

$$f(E) = \beta(1 + \beta)\frac{EU^\beta}{(E + U)^{2+\beta}} \qquad (16)$$

where E is the energy, β the shape parameter (equal to 0.7 for Na) and U the characteristics energy ($\approx 0.05$ eV for Na). A cut-off function (Emax$\approx$100 eV) has been used in order to eliminate the high energy tail of the function, because the maximum ejection energy should be lower than the photon energy peaked at the Lyα line (10.2 eV) in this energy range. A weight w has been assigned to each particle in order to simulate the flux. A cubic 3D accumulation grid is defined, with edges $-12$ RM $< x < 3$ RM, $-2$ RM $< y, z < 2$ RM where RM=2440 km is the radius of Mercury and the x axis points toward the Sun; for each cell the Na density is calculated with the number of test particles, the weight w and the lifetime in the cell (more details in Mura et al. 2007).

**3 Results and discussion**

In **Fig. 1** we show some temperature curves (temperature as function of time normalized to the solar Hermean day, that is, 176 days, equivalent to 2 Hermean years) calculated in the points (longitude-latitude) indicated in the legend. The reference temperature (T $\propto \cos(\alpha)^{1/4}$ where $\alpha$ is the illumination angle) is higher than the thermophysical temperature because it is calculated by assuming instantaneous equilibrium between received and emitted energy (zero thermal inertia); the thermophysical temperature assumes instead heat propagation in the soil (non-zero thermal inertia), so the surface does not reach the temperature it would reach if the whole received energy were adsorbed on the surface. The thermophysical temperature therefore provides a more realistic description. As we can see in **Fig. 1**, the difference between reference and modelled temperature increases towards the poles and at the terminator; this is because the solar irradiance ($\propto \cos(\alpha)$, see eq. 3) decreases when latitude increases, so the heat propagation in the thermophysical model becomes more important with respect to the surface irradiation.

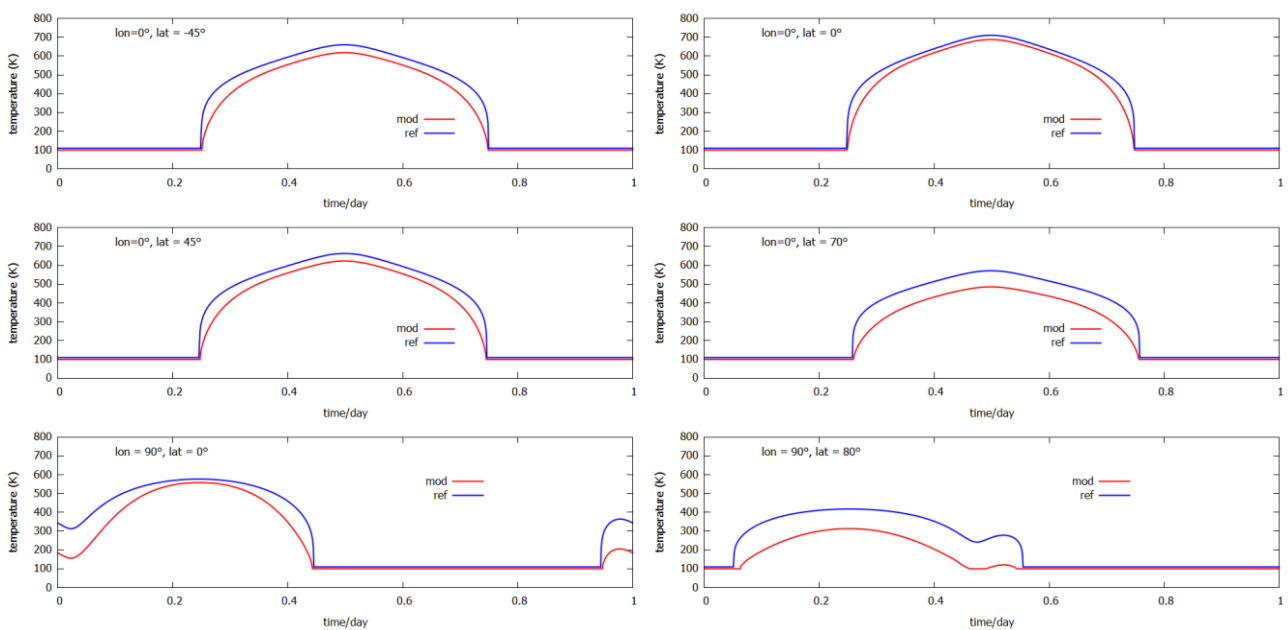

**Fig. 1.** Temperature as function of time (normalized to solar Hermean day, 176 terrestrial days) for different locations (longitude-latitude points), calculated with the thermophysical code (red line) and with the reference temperature law (T proportional to ¼ power of illumination angle cosine, blue line). The local temperature rises (bottom plots) are due to temporary retrograde motion of the Sun, because the planet is at the perihelion and the angular orbital velocity is greater than the spin angular velocity.

The situation is more clearly depicted in **Fig. 2**, where we show the thermal map of the difference between the reference temperature and the temperature calculated with the thermophysical model, at different values of true anomaly angle. Angle greater than 360 degrees refers to the second revolution around the Sun, because the Hermean day is equivalent to two orbital periods. The difference is zero at nighttime, because the two temperatures are set at the same value.

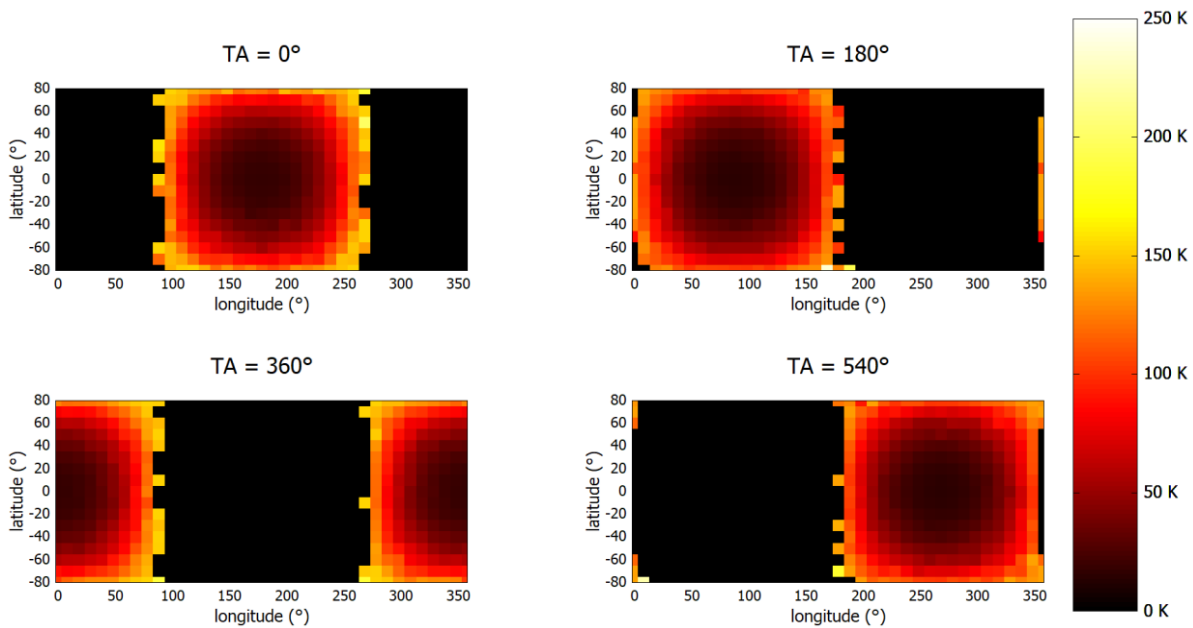

**Fig 2.** Difference between reference temperature and thermophysical temperature, for the true anomaly angles indicated in the legends.

In **Fig. 3,** the comparison of the total numerical content of the Na exosphere as a function of time and true anomaly angle, computed by using the "reference temperature" distribution and the temperature distribution derived with our thermophysical code (nighttime temperature set to 100 K). The model has been calculated by assuming a low thermal inertia type material ("fine dust", $TI \sim 10\ J\ s^{-1/2}\ m^{-2}\ K^{-1}$), according to previous estimates that assign an average low thermal inertia to the planet (see for example Morrison, 1970).

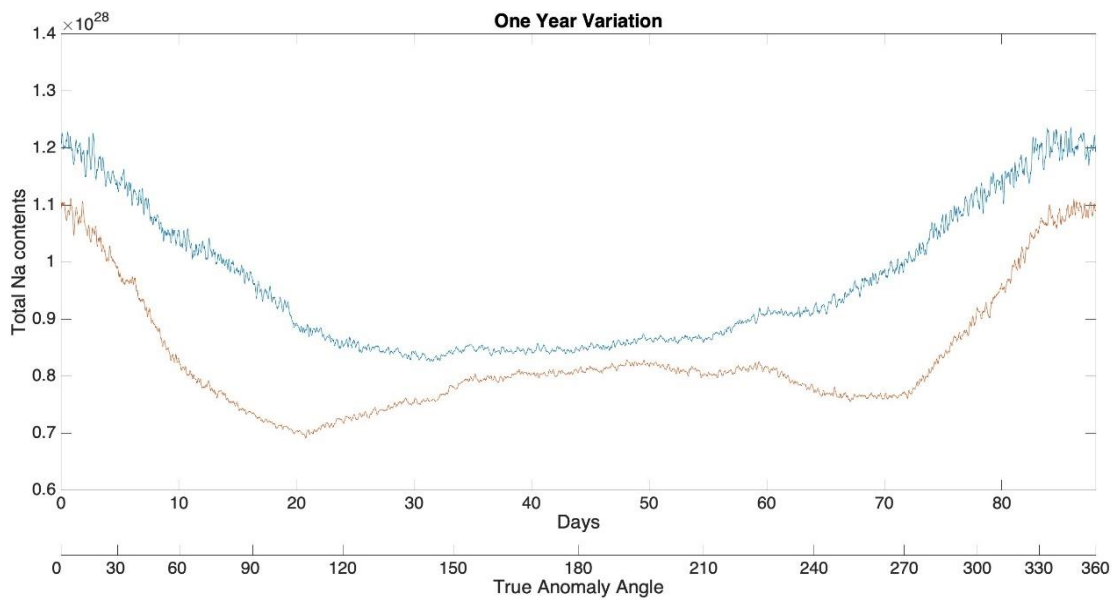

**Fig. 3.** Total sodium exospheric content as function of time and true anomaly angle, calculated in two cases: surface temperature from the thermophysical code (red line), and reference temperature (T proportional to ¼ power of cosine of illumination angle, blue line).

For better clarifying the IAPS model outputs, **Fig. 4** shows the exosphere from the model of Mura et al., 2009, considering as input the reference temperature" distribution (while the nighttime temperature was set to 100 K). Each figure shows the simulated exosphere sodium density, plotted over 3 perpendicular planes and at the lowest altitude (i.e. just above Mercury surface), for a value of true anomaly angle of 90 degrees, and for the two processes of emission considered in the model:

PSD and TD. In the model, the emission of sodium is regulated by its abundance in the uppermost surface layer (which is calculated by the model at each iteration). The model includes the precipitation of plasma (which enhances the diffusion of sodium from the inside to the outside of the regolith grains), the circulation of exospheric sodium, the radiation pressure acceleration (variable with time, and maximum around 60 and 300° of TAA), photoionization of sodium, IS (not shown in the figure), the rotation of the surface (including the short period of retrograde apparent motion close to perihelion), etc. Electron stimulated desorption is presently not included in the model, but it could be easily included, as well. We can see that the sodium particles ejected in the exosphere by the TD have velocities much lower than the escape velocity and they fall down onto the surface; the PSD process, influenced by proton impact onto the surface, is then responsible of the dominant Na population at high altitude.

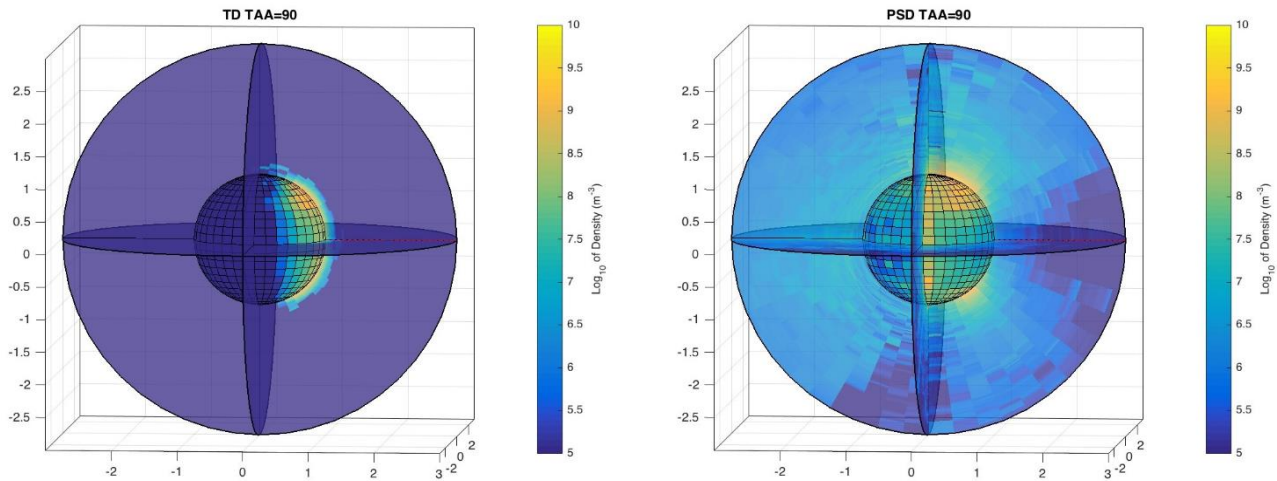

**Fig 4.** Example of results from the model of Mura et al., 2009. The simulated exosphere sodium density is plotted over 3 perpendicular planes and at low altitude for the indicated true anomaly angle (degrees), and for the two processes of emission considered in the model: Photon Stimulated Desorption (PSD) and Thermal Desorption (TD).

In this version of the model, with the used parameters describing diffusion inside the grains, radiation pressure and photoionization effects, the model reproduces an exospheric total content of Na peaking close to the perihelion. In the case of the reference model, we can discriminate only a primary peak at TAA = -30° (Fig. 3). This well reproduces the effect of the temporary retrograde motion of the Sun, also visible in the temperature maps at TAA = 0°. When using the new temperature model, the exosphere shows a similar primary peak at the same TAA, but also a secondary large peak extending to the whole aphelion phase. Since the only difference between the two cases is the input model of the surface temperature, we can state that a more realistic surface temperature model would produce an enhancement of the total Na content close to aphelion.

The sodium abundance calculated with the reference model is higher with respect to those calculated by using the thermophysical model. This reflects the estimated contribution due to TD process. In fact, the TD is highly temperature dependent ($\propto e^{-U/k_B T}$, eq. 15); especially at higher latitudes, where the temperature difference is higher, the thermal desorption is rather different between the reference and thermophysical case. While the TD is temperature dependent, the PSD does not; with a fixed number of test particles launched on the exosphere simulation, if the temperature increases, the sodium particles number emitted for TD increases, while those of PSD decreases. The surface-integrated difference between temperature models on the whole surface could explain the different sodium exospheric content.

As described above the absolute maxima values in sodium content shown in **Fig. 3** correspond to passages close to the perihelion, when the average temperature is at maximum. The secondary peaks correspond instead to passages to the aphelion. However, the real observations show a different trend. In fact, as reported in different studies based on ground based (Potter et al. 2007, Milillo et al 2021) and space (Cassidy et al. 2015) observations, the total content of the Na exosphere have maxima close to perihelion and aphelion but the main and secondary peaks are inverted (i.e. the main peak occurs at the aphelion while the secondary at the perihelion).

The exospheric behavior can be explained by a mix of source, loss and transport processes; indeed the PSD may be responsible of the ejection of most of the sodium exosphere, but no peak near aphelion would be present if the numerical sodium content were only dependent on PSD. The radiative acceleration of a sodium atom is a function of velocity due to the Doppler effect, changing the solar continuum seen by atoms and so their emission. The radiative acceleration is minimum at the perihelion and at the aphelion, thus also the loss due to the acceleration in the anti-solar direction is minimum [Potter et al., 2006]. At the perihelion, because of the short heliocentric distance, the photon flux, following a $r^{-2}$ law, is however high and the photoionization rate is maximum, while at the aphelion the photoionization is minimum and the loss rate is minimized. The aphelion peak of total Na content may also occur because of the high terminator velocity, that causes high ejection of sodium adsorbed on nighttime [Leblanc and Johnson, 2003]. Cassidy et al. (2016), by analyzing the MESSENGER density profiles, identified an extra source for Na surface release possibly due to a preferred condensation of this species where the average surface temperature is minimum (so called cold poles).

The model considers preferential sticking as a function of temperature, so higher at the coldest regions (+90° and -90° lon) also the the same coldest regions are also those that spend much more time in the nighside having much more time for Na accumulation. The Na exosphere increase at aphelion can be explained in the model as the effect of colder regions, rich in "free" Na atoms, that are illuminated by Sun light (approaching aphelion, see Fig. 1 bottom left panel and Fig. 3), the gradual thermal variation reproduced by the thermal model makes the Na release due to TD more gradual than considered before, thus producing the second peak.

**Conclusions**

The link between the surface temperature of Mercury and the total exosphere sodium content has been studied by comparing the results obtained with two different temperature inputs: a "reference temperature" (proportional to ¼ power of illumination angle cosine), and a more realistic temperature calculated with a thermophysical code that takes into account the thermal conductivity of the soil. We found that the exospheric content increases with temperature. The temperature calculated with the more realistic thermophysical code is lower than the reference temperature case because the non-zero thermal inertia is considered; as a consequence, the exospheric sodium densities decrease. The exosphere content is maximum at perihelion because the temperature attains its maximum value and, hence, maximum TD release, and a local maximum is present also at the aphelion when the temperature is lowest; these two maxima have been confirmed by observations (although the main peak is located at the perihelion and not the aphelion), and it can be explained with a mix of source, loss and transport processes. The radiative acceleration rate are minimum at both the apsides and photoionization are maximum at perihelion and minimum at aphelion. At perihelion the terminator velocity is maximum, so the sodium adsorbed on nighttime is quickly ejected; the anti-sunward sodium transport is also minimized because the radiative acceleration is minimum.

Possible, explanations of the peak at aphelion reproduced in the new model result are that the cold poles have higher sticking efficiency and spend longer time in the night side accumulating more sticked Na. Given the smoother thermal variations of the cold poles while approaching the aphelion obtained with the thermal model, the Na atoms are released more gradually (lower contribution due to TD) as the region is in the Sunlight, thus producing a wide increase of Na content in the whole aphelion phase.

To fully explain the Na observation trend along the Mercury's orbit an extra source at the cold poles is required. In the near future we plan to study the effects of surface heterogeneities, taking into account different geological features as smooth plains.

**References**

Acton, C. H. (1996), Ancillary data services of NASA's Navigation and Ancillary Information Facility, *Planetary and Space Science*, **44**, 65-70.

Bida, Thomas A., et al. (2000), Discovery of calcium in Mercury's atmosphere, *Nature*, **404**, 159-161.

Bida, T. A., Killen, R. M. (2017), Observations of the minor species Al and Fe in Mercury's exosphere, *Icarus*, **289**, 227-238.

Biele, J., Pelivan, I., Kuhrt, E., Davidsson, B., Choukroun, M., Alexander, C. J. (2014), Recommended values and correlations of thermophysical properties for comet modelling, American Geophysical Union, Fall Meeting 2014, abstract id.P41C-3915.

Broadfoot, A. L., Kumar, S., et al. (1974), Mercury's Atmosphere from Mariner 10: Preliminary Results, *Science*, **185**, 166-169.

Capria, M. T., Tosi, F., De Sanctis, M. C., Capaccioni, F., et al. (2014), Vesta surface thermal properties map, *Geophysical Research Letters*, **41**, doi:10.1002/2013GL059026

Cassidy, T. A., Merkel, A. W., Burger, M. H., et al. (2015), Mercury's seasonal sodium exosphere: MESSENGER orbital observations, *Icarus*, **248**, 547-559

Cassidy, T. A., McClintock, W. E., Killem, R. M., et al. (2016), A cold-pole enhancement in Mercury's sodium exosphere, *Geophys. Res. Lett.*, **43**, 11, 121–11,128, doi:10.1002/2016GL071071.

Cox, A. N. (2000), Allen's Astrophysical Quantities, 4th edn. (New York: AIP Press, Springer)

Cremers, C. J., Hsia, H. S. (1973), Thermal conductivity of Apollo 15 fines at low density (abstract), *Lunar Science-IV*, 164-166. The Lunar Science Institute, Houston.

Davidsson, B. J. R., Rickman, H., et al. (2015), Interpretation of thermal emission. I. The effect of roughness for spatially resolved atmosphereless bodies, *Icarus*, **252**, 1-21.

ESA Mercury Environmental Specification, BepiColombo definition study, SCI-PF/BC/TN01.

Glaser P., Glaser, D. (2019), Modeling near-surface temperatures of airless bodies with application to the Moon, *Astron. Astrophys.*, **627**, 1-14.

Grava C., Killen R.M., et al. (2021), Volatiles and refractories in surface-bounded exospheres in the inner Solar System, *Space Science Reviews*, **217**, article id. 61.

Hayne, P. O., Bandfield, J. L., et al. (2017), Global regolith thermophysical properties of the Moon from the Diviner Lunar Radiometer Experiment, *Journal of Geophysical Research: Planets*, **122**, 2371–2400. https://doi.org/10.1002/2017JE005387.

Hunten, D.M., Morgan, T.H., Shemansky, D.E. (1988), The Mercury atmosphere. In: Vilas, F., Chapman, C.R., Matthews, M.S. (Eds.). Mercury. Univ. of Arizona Press, Tucson, pp. 613–621.

Johnson, R.E., Leblanc, F., Yakshinskiy, B.V., Madey, T.E. (2002), Energy distributions for desorption of sodium and potassium from ice: The Na/K ratio at Europa, *Icarus*, **156**, 136–142.

Keihm, S. J., et al. (2012), Interpretation of combined infrared, submillimeter, and millimeter thermal flux data obtained during the Rosetta fly-by of Asteroid (21) Lutetia, *Icarus*, **221**, 395–404.

Killen, R.M., Potter, A.E., et al. (2001), Evidence for space weather at Mercury, *Journal of Geophysical Research*, **106**, 20509–20526.

Killen, R. M., Sarantos, M., et al. (2004), Source rates and ion recycling rates for Na and K in Mercury's atmosphere, *Icarus*, **171**, 1-19.

Killen, R., C., G., et al. (2007), Processes that Promote and Deplete the Exosphere of Mercury, *Space Science Reviews*, **132**, 433-509.

Leblanc, F., Johnson, R.E., 2003, Mercury's sodium exosphere, *Icarus,* **164**, 261–281.

Leblanc, F., Doressoundiram, A., et al. (2009), Short-term variations of Mercury's Na exosphere observed with very high spectral resolution, *Geophysical Research Letters*, **36**, CiteID L07201.

Leblanc, F., Johnson, R. E. (2010), Mercury exosphere I. Global circulation model of its sodium component, *Icarus*, **209**, 280-300.

LeBlanc, F., Doressoundiram, A. (2011), Mercury exosphere. II. The sodium/potassium ratio, *Icarus*, **211**, 10-20.

Mangano, V., Massetti, S., et al. (2015), THEMIS Na exosphere observations of Mercury and their correlation with in-situ magnetic field measurements by MESSENGER, Planetary and Space *Science*, **115**, 102-109.

Massetti, S., Mangano, V., et al. (2017), Short-term observations of double-peaked Na emission from Mercury's exosphere, *Geophysical Research Letters*, **44**, 2970-2977.

McClintock, W. E., Bradley, E. T., et al. (2008), Exploring Mercury's Surface-Bound Exosphere with the Mercury Atmospheric and Surface Composition Spectrometer: AN Overview of Observations during the First Messenger Flyby, *American Geophysical Union*, Spring Meeting 2008, abstract id.U24A-02.

Milillo, A., Fujimoto, M., et al. (2010), The BepiColombo mission: An outstanding tool for investigatingthe Hermean environment, *Planetary and Space Science*, **58**, 40-60.

Milillo, A., Fujimoto, M., et al. (2020), Investigating Mercury's Environment with the Two-Spacecraft BepiColombo Mission, *Space Sci. Rev.*, **216**, 93.

Milillo, A., Mangano, V., et al. (2021), Exospheric Na distributions along the Mercury orbit with the THEMIS telescope, *Icarus*, **355**, 114179.

Morrison, D. (1970), Thermophysics of the planet Mercury, *Space Sci. Rev.*, **11**, 271–307.

Müller, T. G., and J. S. V. Lagerros (1998), Asteroids as far-infrared photometric standards for ISOPHOT, *Astron. Astrophys.*, **338**, 340–352.

Mura, A., Orsini, S., Milillo, A., Delcourt, D., Massetti, S. (2005), Dayside H+ circulation at Mercury and neutral particle emission, *Icarus*, **175**, 305–319.

Mura, A., Milillo, A., Orsini, S., Massetti, S. (2007), Numerical and analytical model of Mercury's exosphere: Dependence on surface and external conditions, *Planetary and Space Science* **55** (11), 1569–1583.

Mura, A., Wurz, P., Lichtenegger, H. I. M., et al. (2009), The sodium exosphere of Mercury: Comparison between observations during Mercury's transit and model results, *Icarus*, **200**, 1-11

Mura, A. (2012), Loss rates and time scales for sodium at Mercury, *Planetary and Space Science*, **63**, 2-7.

Orsini, S., Livi, S., et al. (2021), SERENA: Particle Instrument Suite for Determining the Sun-Mercury Interaction from BepiColombo, *Space Science Reviews*, **217**, 1.Potter, A. E., Killen, R. M., Morgan, T. H. (1999), Rapid changes in the sodium exosphere of Mercury, *Planetary and Space Science*, **47**, 1441-1448.

Potter, A., Killen, R.M., Morgan, T.H. (2002), The sodium tail of Mercury, *Meteor. Planet. Sci.*, **37** (9), 1165–1172.

Potter, A. E., Killem, R. M., Sarantos, M. (2006), Spatial distribution of sodium on Mercury, *Icarus*, **181**, 1-12.

Potter, A. E., Killen, R. M., Morgan, T. H. (2007), Solar radiation acceleration effects on Mercury sodium emission, *Icarus*, **186**, 571-580.

Potter, A. E., Morgan, T. H. (1985), Observations of Sodium on Mercury, *Bulletin of the American Astronomical Society*, **17**, 710

Potter, A. E., Morgan, T. H. (1986), Potassium in the atmosphere of Mercury, *Icarus*, **67**, 336-340.

Rognini, E., Capria, M. T., Tosi, F., De Sanctis, M. C., et al. (2019), High Thermal Inertia Zones on Ceres From Dawn Data, *Journal of Geophysical Research*, https://doi.org/10.1029/2018JE005733.

Sakatani, N., Ogawa, K., et al. (2017), Thermal conductivity model for powdered materials under vacuum based on experimental studies, *AIP Advances,* **7**, 015310. Doi: 10.1063/1.4975153.

Sarantos, M., Killen, et al. (2008), Correlation between lunar prospector measurements and the lunar exosphere during passage through the Earth's magnetosphere, *Geophys. Res. Lett.*, **35**. L04105.

Schleicher, H., Wiedemann, G., et al. (2004), Detection of neutral sodium above Mercury during the transit on 2003 May 7., *Astronomy and Astrophysics*, **425**, 1119–1124.

Slavin, M. H. Acuña, Anderson, B. J., et al. (2008), Mercury's magnetosphere after MESSENGER's first flyby, *Science*, **321**, 85-89.

Smyth, W.H., Marconi, M.L. (1995), Theoretical overview and modeling of the sodium and potassium atmospheres of Mercury, *Astrophysical Journal*, **441**, 839–864.

Vasavada, A. R., Paige, D. A. (1999), Near-Surface Temperatures on Mercury and the Moon and the Stability of Polar Ice Deposits, *Icarus*, **141**, 173-193.

Vasavada, A. R. et al. (2012), Lunar equatorial surface temperatures and regolith properties from the Diviner Lunar Radiometer Experiment, *Journal of Geophysical Research*, **117**, E00H18.

Yakshinskiy, B.V., Madey, T.E. (1999), Photon-stimulated desorption as a substantial source of sodium in the lunar atmosphere, *Nature*, **400**, 642–644.

Yakshinskiy, B.V., Madey, T.E., Ageev, V.N. (2000), Thermal desorption of sodium atoms from thin SiO2 films, *Surface Rev. Lett*., **7**, 75–87.